\documentclass[11pt,twoside]{article}
\usepackage{asp2010}
\usepackage{graphicx} 
\usepackage{natbib,url} 
\resetcounters

\begin{document}

\title{Reflection and refraction of (magneto-)acoustic waves at the magnetic canopy: further evidences from multi-height 
seismic data}
\author{S.P. Rajaguru,$^1$ Xudong Sun,$^2$ K. Hayashi, $^2$ and S. Couvidat, $^2$
\affil{$^1$ Indian Institute of Astrophysics, Koramangala II Block, Bangalaore, India}
\affil{$^2$ Hansen Experimental Physics Laboratory, Stanford University, Stanford CA 94305-4085, USA}}

\begin{abstract}
We recently presented evidences \citep{rajaguruetal12} that seismic halos around expanding 
magnetic structures in the lower solar atmosphere are related to the acoustic to magnetoacoustic wave conversions,
using multi-height data from Helioseismic and Magnetic
Imager (HMI) and Atmospheric Imaging Assembly (AIA) (1700 and 1600 \AA~ channels) onboard Solar Dynamics Observatory (SDO). 
Using the same data, we here present and discuss further evidences through analyses of maps of phase-shifts
between observables from different heights and their correspondence with oscillation power.
The phase shift maps provide more direct signatures of reflection and refraction of (magneto-)acoustic wave modes.

\end{abstract}

\section{Introduction}

A rich variety of wave-dynamical phenomena in and around structured magnetic field, large and small, that thread the
solar atmospheric layers, are now amenable to detailed observational studies thanks to a variety of instruments,
both space- and ground-based, that offer multi-line and mult-height measurements [see the review and references compiled by \citet{khomenko13}]. 
Acoustic or seismic halos are regions of enhanced power, when compared to the quiet-Sun,
surrounding strong-magnetic-field structures such as sunspots and plages at frequencies
above the photospheric acoustic cut-off of $\approx$ 5.3 mHz, in the range of 5.5 -- 7 mHz, and over
regions of weak to intermediate strength (50 -- 250 G) photospheric magnetic field \citep{1992ApJ...394L..65B,1992ApJ...392..739B,1998ApJ...504.1029H}
(see \citet{rajaguruetal12} for further references and summary of existing observational and theoretical work).
In our above referred recent work, through detailed analyses of power maps derived from photospheric Doppler velocity, continuum and line core intensities observed by 
the \textit{Helioseismic and Magnetic Imager} [HMI:\citet{2012SoPh..275..207S}] and 
that from upper photospheric and lower chromospheric UV emissions in the 1700 \AA~  and 1600 \AA~ wavelength
channels imaged by the \textit{Atmospheric Imaging Assembly} [(AIA: \citet{2012SoPh..275...17L}] onboard the \textit{Solar Dynamics Observatory} (SDO), 
we presented evidences that the seismic halos 
are related to the acoustic to magnetoacoustic wave conversions \citep{2009A&A...506L...5K}.
Here, using the same data sets, we analyse maps of phase-shifts
between observables from different heights and their spatial correspondence with power maps as well
as inclination and strength of magnetic field derived from HMI vector field observations.

\section{Data and Analysis Method}
The data sets used are the same as that in \citet{rajaguruetal12} (we refer the readers to this paper for details):
14-hour-long tracked data cubes of photospheric Doppler velocity [$v$], continuum intensity [$I_{\rm{c}}$],
line-core intensity [$I_{\rm{co}}$], and dis-ambiguated vector magnetic field [$B_{\rm{x}}, B_{\rm{y}},$ and $B_{\rm{z}}$] derived from
HMI observations, and chromospheric UV emissions observed by AIA in the wavelength channels 1700 \AA~ and 1600 \AA, which
we denote as $I_{\rm{uv1}}$ and $I_{\rm{uv2}}$, respectively. The spatial extent of the region covered is 512 $\times$ 512 pixels,
with a sampling rate of 0.06 degrees (heliographic) per pixel.
In this paper we analyse the sunspot region NOAA 11092, whose central meridian passage was on 3 August 2010.

In addition to the power maps calculated as described in \citet{rajaguruetal12}, we derive phase difference maps between the same physical
variables. 
We analyse the spatial correspondence between power and phase maps through azimuthal averages (centered around the sunspot), and their
relation to similarly averaged magnetic variables, total field strength and inclination.

\section{Power and phase maps}

The $I_{\rm{c}}$, $I_{\rm{co}}$, and $v$ from HMI form at three different heights spread over
$z$ = 0 -- 300 km above the continuum optical depth $\tau_{\rm{c}}$ = 1 ($z$ = 0 km) level:
$I_{\rm{c}}$ is from about $z$ = 0 km, $v$ corresponds to an average height
of about $z$ = 140 km \citep{2011SoPh..271...27F}, and the line core intensity $I_{\rm{co}}$ corresponds
to the top layer, at about $z$ = 280 -- 300 km, of the line formation region \citep{2006SoPh..239...69N}.
The AIA 1700 \AA~ and 1600 \AA~ intensities
[$I_{\rm{uv1}}$ and $I_{\rm{uv2}}$], form at average heights of 360 km and 430 km \citep{2005ApJ...625..556F}, respectively.
We have calculated phase shifts between all the intensities, $\phi_{\rm{I-I}}$, but for the purpose of this short 
presentation we present and discuss only those between $I_{\rm{c}}$ and [$I_{\rm{co}}$, 
$I_{\rm{uv1}}$].
Figure 1 displays the spatial maps of phase-shifts $\phi_{\rm{I_{c}-I_{co}}}$ ({\it{left panel}}), and 
of phase shifts $\phi_{\rm{I_{c}-I_{uv1}}}$ ({\it{right panel}}). Comparison of these phase shifts with the corresponding
power maps from HMI Doppler $v$ and AIA $I_{uv1}$ in Figures 2 and 4 of \citet{rajaguruetal12} shows that
regions of enhanced power correspond to a nulling of or 
much reduced phase shifts implying both upward and downward propagation in the layers between $z=$ 0 to 300 km. All over the 
non-magnetic quiet region there is only upward propagation as evidenced by the uniform positive phase shifts: while for
$\phi_{\rm{I_{c}-I_{co}}}$  there is gradual increase in the phase shifts as frequency $\nu$ increases, $\phi_{\rm{I_{c}-I_{uv1}}}$
display decreased values again at $\nu$= 8 mHz implying that such frequency waves begin to be evansecent (i.e., the cut-off frequency
is reached) at heights where $I_{uv1}$ forms.  
\begin{figure}[!ht]
\centerline{\hspace*{0.10\textwidth}
               \includegraphics[width=0.9\textwidth,height=0.45\textheight,clip=]{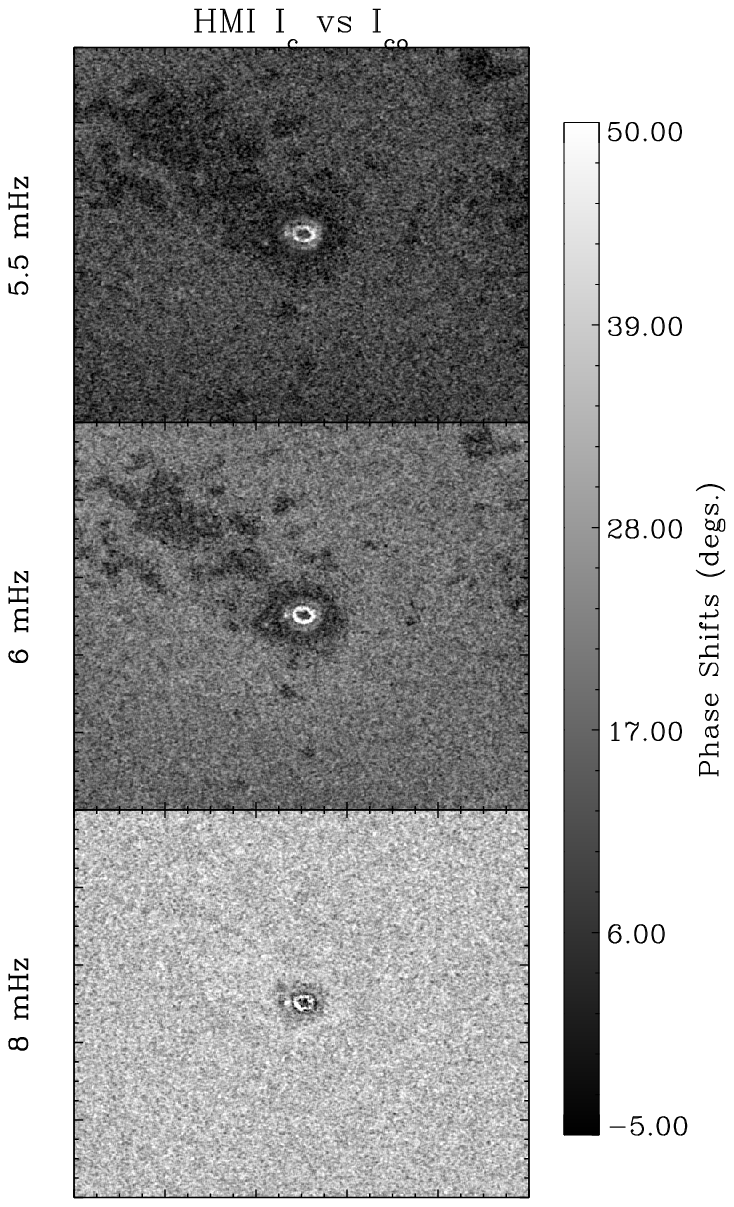}
               \hspace*{-0.50\textwidth}
               \includegraphics[width=0.9\textwidth,height=0.45\textheight,clip=]{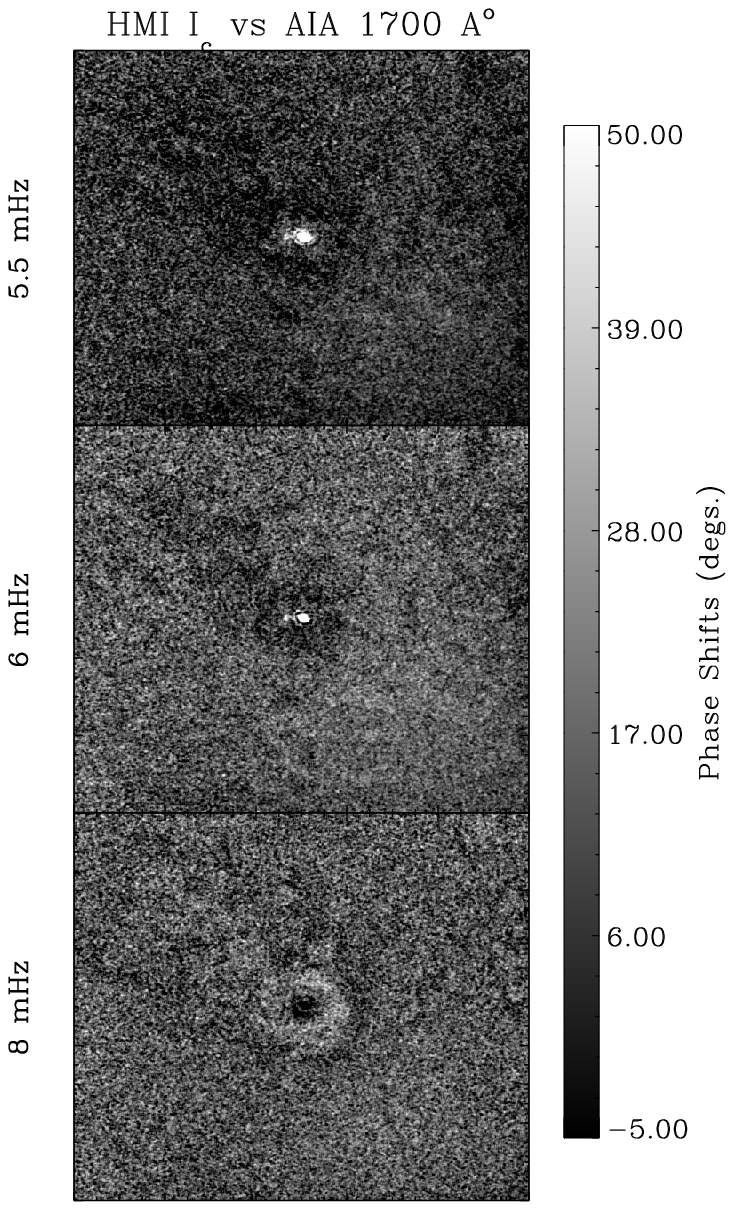}
              }
     \vspace{-0.986\textwidth}   
     \vspace{0.93\textwidth}    
\caption{Maps of phase-shifts $\phi_{\rm{I_{c}-I_{co}}}$ ({\it{left panel}}), and                              
of phase shifts $\phi_{\rm{I_{c}-I_{uv1}}}$ ({\rm{right panel}}) over the sunspot region NOAA 11092 for three different frequencies as marked. The region shown here
covers a square area of 373 $\times$ 373 Mm$^{2}$.}
\label{fig1}
\end{figure}
A closer examination of the spatial correspondence between power and phase variations is done through spot-centered azimuthal
averages of power and phase shifts. These quantities are plotted in Figure 2 along with (in the bottom panels) similarly
averaged magnetic quantities, total magnetic field strength $B_{\rm{tot}}$ and inclination $\gamma$ with respect to the vertical direction.
The {\it{left panel}} is for 6.0 mHz, and the {\it{right panel}} is for 8.0 mHz; solid curves show the phase shifts, while the 
dotted curves show the power normalised to quiet-Sun values. The horizontal dashed lines give the quiet-Sun average phase shifts,
the horizontal dotted lines are at the normalised power level 1 (i.e. quiet-Sun level), and the two vertical dot-dashed lines
mark the umbral (6.6 Mm) and outer-penumbral (17.5 Mm) radii. For heights less than about 300 km, as seen in the top two
panels of Figure 2, there is a close correspondence between nulled or much reduced (as compared to quiet-Sun levels) phase
shifts and the acoustic halos (or enhanced power) implying contributions from reflected or downward (following refraction)
propagation of compressive waves. However, for acoustic halos at heights where AIA 1700 \AA~($I_{\rm{uv1}}$) form, there seem
to be larger contributions from upward propagating waves.

\begin{figure}[!ht]
\centerline{\hspace*{0.1\textwidth}
               \includegraphics[width=0.82\textwidth,height=0.45\textheight,clip=]{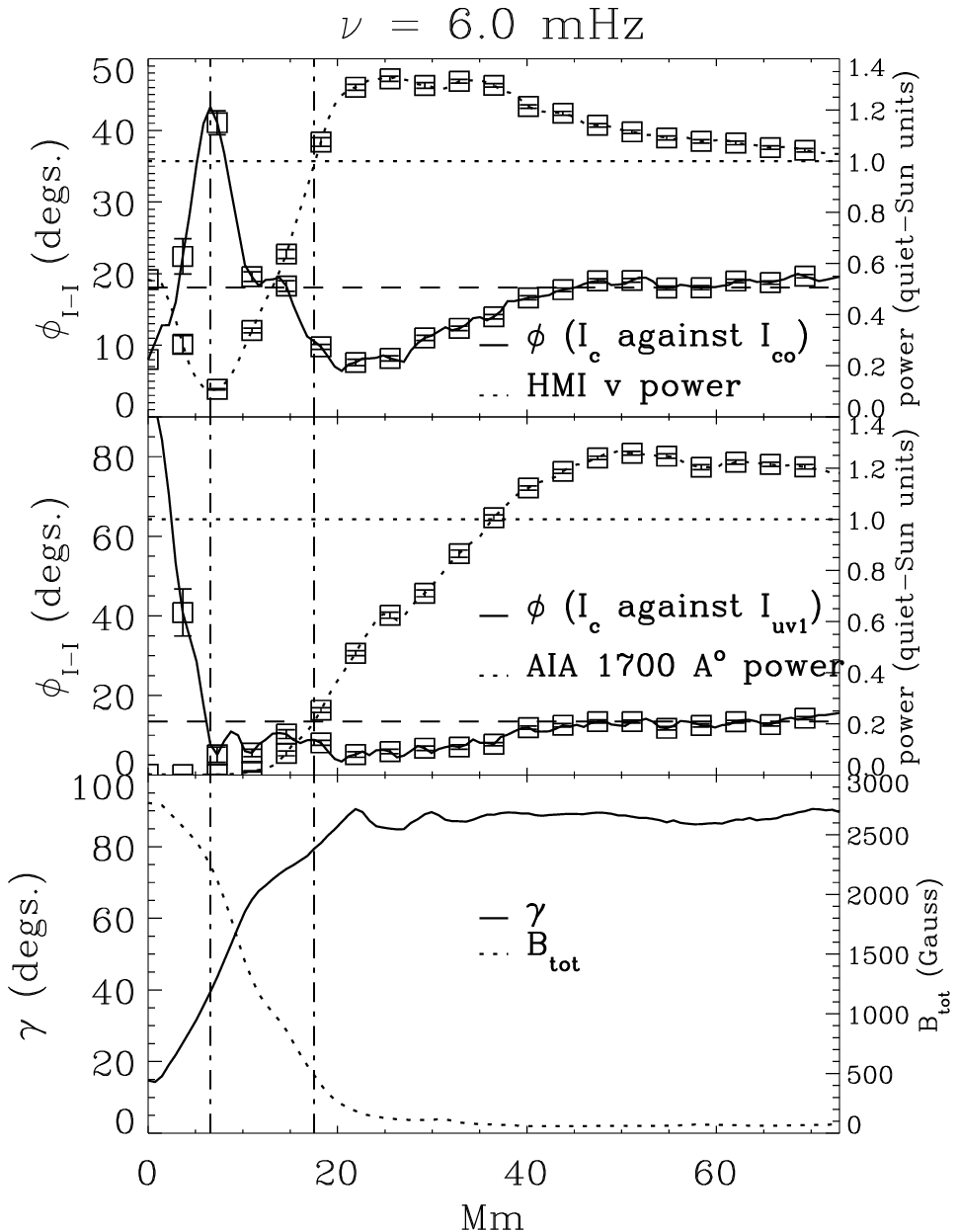}
               \hspace*{-0.35\textwidth}
               \includegraphics[width=0.82\textwidth,height=0.45\textheight,clip=]{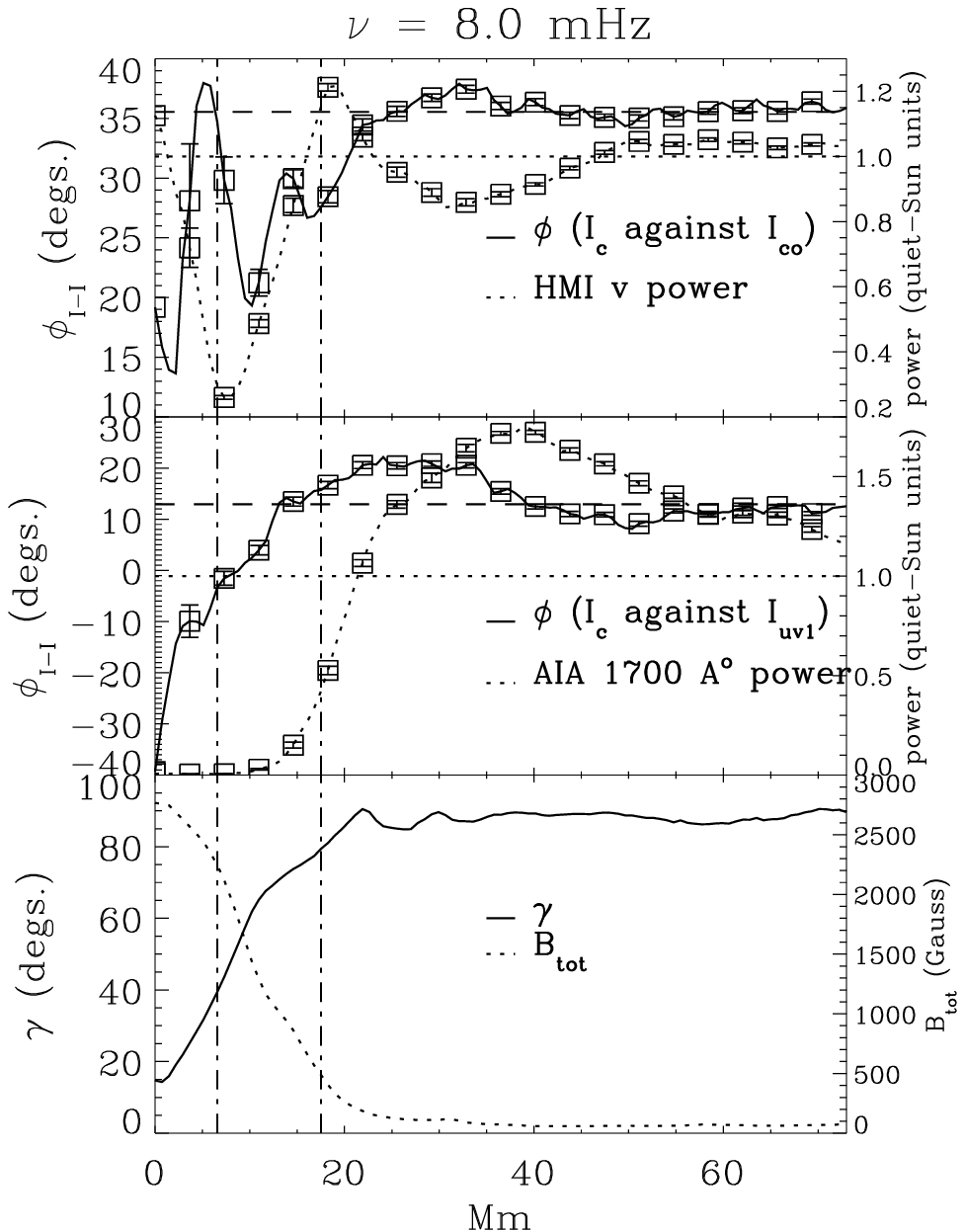}
              }
     \vspace{-0.74\textwidth}   
     \vspace{0.75\textwidth}    
\caption{Azimuthal averages of normalised power and phase shift maps calculated between observables as marked in the panels. The
bottom panels show the similarly averaged total magnetic field strength $B_{\rm{tot}}$ and inclination $\gamma$. 
Error bars depict the standard errors of the means plotted. Refer to the text for further details.}
\label{fig3}
\end{figure}

Another interesting behaviour of phase and power variation is seen in the umbral and umbral-penumbral boundary region:
above 5 mHz (here in the figure at 6 mHz) there is enhanced propagation between the height levels of $I_{\rm{c}}$ and $I_{\rm{co}}$
coinciding with highly suppressed power as seen in the intermediate level of Doppler $v$, while at higher levels this enhanced
propagation 'migrates' to the the umbra and is the dominant propagation signal at 6 mHz as seen in the middle panel.
It is to be noted that inclined $B$ fields of penumbrae harbor propagating waves even at $\nu$ well below the photospheric
cut-off of $\approx$ 5.2 mHz [not shown here; but have been studied in detail, e.g. in \citet{2010ApJ...721L..86R}]. 
However, at 8 mHz, there is a large negative phase shift showing downward propagation. We see this to be a possible result
of refraction and reflection, above the umbral layers at heights above the formation heghts of $I_{\rm{uv1}}$, of waves
of 8 mHz. In fact, this downward propagation is observed to start at about 7 mHz (not shown in the figures).

More detailed analyses, including further results on phase shifts between other observables not shown and discussed here, are
being prepared for a publication in a peer-reviewed journal.

\acknowledgements Xudong Sun, K. Hayashi, and S. Couvidat are supported by NASA grants NNG05GH14G to the SDO/HMI project
at Stanford University. The data used here are courtesy of NASA/SDO and the HMI and AIA science teams.

\bibliography{ms}

\end{document}